\documentstyle[12pt,axodraw]{article}
\begin{document}
\begin{flushright}
{DESY 99-182\\
December 1999}
\end{flushright}
\vspace{1.5cm}

\begin{center}
{\Large \bf Probe of the $Wtb$ coupling in $t \bar{t}$ pair production at Linear
Colliders\\
\hfill\\}
\end{center}
\vspace{0.5cm}

\begin{center}
{E.~Boos$^{1,2}$, M.~Dubinin$^{1,2}$,
M.~Sachwitz$^2$, H.J.~Schreiber$^2$ \\ 
\hfill\\
{\small \it $^1$Institute of Nuclear Physics, Moscow State University}\\
{\small \it 119899, Moscow, Russia} \\
{\small \it $^2$DESY-Zeuthen, D-15738 Zeuthen,
Germany}}
\end{center}

\vspace{1.0cm}
\begin{center}
{\bf Abstract}
\end{center}
\begin{quote}
The $Wtb$ vertex can be probed on future 
colliders in the processes of single top
production (LHC, $pp$ mode, NLC, $\gamma e$ mode)
and of top pair production (NLC, $e^+ e^-$ mode).
We analyse observables sensitive to anomalous $Wtb$
couplings in the top pair production process of $e^+ e^-$ collisions.
In particular, forward-backward and spin-spin asymmetries
of the top decay products and the asymmetry
of the lepton energy spectum are considered.
Possible bounds on anomalous couplings obtained
are competitive
to those expected from the upgraded Tevatron and LHC.
The validity of the infinitely small width
approximation for the three-body top decay is also studied in detail.
\end{quote}

\newpage


\section{Introduction}
One of the primary tasks for the forthcoming hadronic and leptonic 
colliders is a detailed study of the top quark properties, in particular,   
the measurements of the top couplings to gauge fields. Special
interest to such measurements is based on the huge difference of
the top quark mass and all other fermion masses,
providing enhanced
expectations for a signal of new physics at the top
mass scale \cite{peccei}.

Among the top couplings to other particles the $Wtb$ coupling
plays a crucial role because it is responsible for
practically all top quark decays. Therefore the spacetime structure of the
$Wtb$ vertex defines the top total width and the characteristics
of its decay products.

There are two general possibilities to probe and measure directly 
the $Wtb$ vertex structure in collider experiments, either from top 
pair production processes or from reactions of single top
production.
The rate of single top production processes is directly proportional to
the $Wtb$
coupling, and thus it is potentially very sensitive to the
$Wtb$ structure. This was indeed demonstrated
in high energy $\gamma e$ collisions  
\cite{boos1, cao} as well as for the upgraded Tevatron and the LHC
\cite{boos2}.
However, the rate of single top production is usually less than
the top pair production rate, in both the lepton and hadron
colliders.  
On the other hand, the reaction
$e^+ e^- \to t \bar{t} \to W^+ b W^- \bar b$
includes the $Wtb$ coupling only in the subsequent top
decays, with the $t(\bar t)$ on-shell decay rate given,
apart from small finite width corrections, by
the top decay branching fraction to $Wb$, which is
close to 100\%. Consequently, the total
rate depends only negligibly on the $Wtb$ vertex
structure \cite{boos1} and more sensitive
observables, like $C$ and $P$ asymmetries, top polarization and
spin correlations, have to be analysed.

The paper starts with the analysis of the process
$e^+ e^- \to t \bar t \to W^+ b W^- \bar b$
in the infinitely small width
approximation, including anomalous couplings in the $Wtb$ vertex.
The narrow-width approximation
enables qualitative interpretations of precise calculations
presented later in this study. In Sect. 4 we perform precise tree-level
computations in the Standard Model (SM) and in the generalization with the
effective $Wtb$ vertex. Asymmetries, energy distributions and
spin-spin correlations are studied, including the option of electron
beam polarization. In Sect. 5 the bounds of the anomalous
coupling parameter space, within those no distinction from the SM
is possible,
are presented and compared
with the corresponding limits obtained from
single top production processes at
NLC in the $\gamma e$ mode and LHC in the $pp$ mode. 

\section{Effective $Wtb$ lagrangian and the anomalous couplings
$f_{2L}$, $f_{2R}$}

In the effective lagrangian approach seven gauge invariant and $CP$ parity
conserving operators of dimension six  \cite{buchmueller-hagiwara-gounaris1,
whisnant} contribute to the $Wtb$ vertex with four independent
formfactors. In our analysis we use the
effective lagrangian in the unitary gauge as given in \cite{boos1,
boos2, KLY} 

\begin{eqnarray}
{\cal L} & \left.=  \frac{g}{\sqrt{2}}\right[ &
 W_{\mu}^-\bar{b}(\gamma_{\mu}f_{1L} P_- +
\gamma_{\mu} f_{1R} P_+) t \nonumber \\
 & & - \left.\frac{1}{2M_W} W_{\mu\nu}
\bar{b}\sigma^{\mu\nu}(f_{2R} P_- + f_{2L} P_+) t
\right] + {\rm h.c.}
\label{eq:lagrangian_anom}
\end{eqnarray}
where $W_{\mu\nu} =
D_{\mu}W_{\nu} - D_{\nu}W_{\mu}, D_{\mu} = \partial_{\mu} - i e A_{\mu},
P_{\pm} = 1/2(1 \pm \gamma_5)$ and \mbox{$\sigma^{\mu\nu}
= i/2(\gamma_{\mu}\gamma_{\nu} - \gamma_{\nu}\gamma_{\mu})$}.

In the SM, the coupling $f_{1L}$ is equal to one
and the other three couplings, $f_{1R}$, $f_{2L}$ and $f_{2R}$, 
are equal to zero. The possible (V+A) coupling $f_{1R}$ is
severely constraint to zero
by the CLEO $b \rightarrow s \gamma$ data \cite{cleo}
on a level \cite{whisnant, larios} which is stronger than expected
even at high energy $\gamma e$ colliders. So in the following,
we set $f_{1R}=$0 and
$f_{1L}=$1 due to the fact that the (V-A) coupling is as
in the SM with the coupling $V_{tb}$ very
close to unity, as required by present data \cite{caso}.
This leaves us to
perform the analysis only for the two 'magnetic' anomalous couplings
$f_{2L}$ and $f_{2R}$.

The couplings $f_{2L}$ and $f_{2R}$ are related to the effective
couplings $C_{tW \Phi}$ and
$C_{bW \Phi}$ \cite{whisnant} in the general effective lagrangian by

\begin{eqnarray}
f_{2L(R)} = \frac{C_{t(b)W \Phi}}{\Lambda^2} \frac{v\sqrt{2} \, m_W}{g}
\end{eqnarray}
where $\Lambda$ is the scale of new physics.
Natural values for couplings $|f_{2L(R)}|$ are of the order 
${\sqrt{m_b m_t}}/{v} \sim$0.1 \cite{peccei}.
Unitarity limit from $t \bar t$ scattering at the scale $\Lambda=$1 TeV
gives the restriction $|C_{tW \Phi}| \le$13.5 \cite{whisnant},
or $|f_{2L(R)}| \le$0.65.
Expected upgraded Tevatron limits on $|C_{tW \Phi}|/(\Lambda/TeV)^2$ are
$\sim$2.6 \cite{whisnant}, so the corresponding upper bounds on
$|f_{2L(R)}|$ are of the order of 0.1-0.2 \cite{boos2}. 

In all our calculations which follow the Feynman rules in
the momentum space corresponding to the effective lagrangian (1) 
were implemented in the program package CompHEP \cite{CompHEP}.

\section{Parity violating observables in the \\
 top decay}

It is straightforward to demonstrate by direct calculation that,
as mentioned in the introduction,
the total rate of the process $e^+ e^- \to t \mu^- \bar \nu_{\mu} \bar b$
is weakly dependent on the anomalous couplings $f_{2L}$ and $f_{2R}$.
For instance, if $(f_{2L}, f_{2R})$= (-0.6, 0), the total cross section
at $\sqrt{s}$= 500 GeV equals 62.7 fb, while the SM value is
63.0 fb. The effect of non-zero $f_{2L,R}$ couplings in the amplitude is
largely compensated by
the increase of the top quark width ($\Gamma_{top}$=1.60 GeV in the
standard case and 4.35 GeV at $(f_{2L},f_{2R})$=(-0.6, 0)). Hence, the
observation of nonstandard interactions is only possible
in variables which are sensitive to the effective
lagrangian terms (1). It is however {\it a priori} not evident
which variables provide sufficiently high sensitivity to anomalous
$Wtb$ operators, so that we are prompted
to look, as a first example, for
the forward-backward asymmetry of top decay products
which is the ratio of integrated single differential distributions.

\subsection{Forward-backward asymmetry in the infinitely
small width approximation}

In the usual approach to the reaction $e^+ e^- \to t \bar t \to$
6 fermions, the final
state topology is calculated in the approximation of infinitely small
top and $W$ widths
\begin{equation}
\frac{1}{(q^2-m^2)^2+m^2\,
\Gamma^2}=\frac{1}{m\, \Gamma} \delta(q^2-m^2) \hspace{0.3cm}.
\end{equation}
Representations of the general expression for distributions in the
$W^+ W^- b \bar b$ final state in terms of the unpolarized $t\bar t$
cross section $\Sigma_{unpol}$, factorized top-antitop branching ratios,
polarization functions $P$, ${\bar P}$ of the $t$, ${\bar t}$ and
the $t \bar t$ spin-spin correlation function $Q$ can be found in
\cite{CDDKZ,BCK}, see also \cite{KRZ}. They can be obtained from the
convolution of the $t\bar t$ production amplitude with the amplitude
density matrices of the $t\to W^+b$ and $\bar t \to W^- b$ decays.
Following the notations of \cite{CDDKZ} one gets
\begin{eqnarray}
\lefteqn{\frac{d^4\sigma(e^+ e^- \to t\bar t\to W^+ b W^- \bar b)} 
 {d\cos \Theta \,d\cos\theta d\varphi \,d\cos\theta^* d\varphi^*}}  \\
\nonumber
  &= \frac{3 \alpha^2 \beta}{32\, \pi\, s} Br(t\to W^+ b)
    Br(\bar t\to W^- \bar b) \Sigma(\Theta, \theta, \varphi,
\theta^*, \varphi^*)& \hspace{0.3cm},
\end{eqnarray}
where $\Theta$ is the top production angle, $\beta=\sqrt{1-4m^2_t/s}$ and
\begin{eqnarray}
\Sigma(\theta, \varphi, \theta^*,
\varphi^*)& =& \Sigma_{unpol}+k\,P\,\cos\theta+\bar k\, \bar P\, \cos
    \theta^* + \cos\theta \cos\theta^* k\,  \bar k \, Q \\ \nonumber
   & & + \quad ( \varphi, \varphi^* \; dependent \; terms).
\end{eqnarray}
The angles $\theta,\,\varphi$/$ \theta^*, \varphi^*$ define the $W$
momentum direction in the rest frame of the top/antitop. The
definitions of these angles can be found in the Appendix. In the 
following, integrations over the azimutal angles $\varphi, \, \varphi^*$
will be always carried out, with the result that $\varphi, \varphi^*$
dependent terms are equal to zero. The variables $k$ and $\bar k$ are
the polarization degree of the top and antitop decay amplitudes.
The expressions for $\Sigma_{unpol}$, $P$, $\bar P$ and $Q$ in terms
of the
helicity amplitudes $\langle \sigma; h_t \, h_{\bar t} \rangle$
for $t \bar t$ production have the form
\begin{eqnarray}
\Sigma_{unpol}=\frac{1}{4}\int d\cos \Theta \sum_{\sigma=\pm}
\Bigl[|\langle \sigma;++\rangle|^2+|\langle\sigma;+-\rangle |^2
+|\langle\sigma;-+\rangle|^2+|\langle\sigma;--\rangle|^2\Bigr] \\
P=\frac{1}{4}\int d\cos \Theta \sum_{\sigma=\pm}
\Bigl[|\langle \sigma;++\rangle|^2+|\langle\sigma;+-\rangle |^2
-|\langle\sigma;-+\rangle|^2-|\langle\sigma;--\rangle|^2\Bigr] \\
\bar P=\frac{1}{4}\int d\cos \Theta \sum_{\sigma=\pm}
\Bigl[|\langle \sigma;++\rangle|^2-|\langle\sigma;+-\rangle |^2    
+|\langle\sigma;-+\rangle|^2-|\langle\sigma;--\rangle|^2\Bigr] \\  
Q=\frac{1}{4}\int d\cos \Theta \sum_{\sigma=\pm}
\Bigl[|\langle \sigma;++\rangle|^2-|\langle\sigma;+-\rangle |^2
-|\langle\sigma;-+\rangle|^2+|\langle\sigma;--\rangle|^2\Bigr]
\end{eqnarray}
where (see, for instance, \cite{BCK})
\begin{eqnarray}
\langle - \mp\pm \rangle&=&\mp (v_L\mp \beta a_L)(1\pm \cos\Theta)\\
\langle - \mp\mp \rangle&=&\pm \frac{2m_t}{\sqrt{s}}v_L \sin\Theta \\
\langle + \mp\pm \rangle&=&\pm (v_R\mp \beta a_R)(1\mp \cos\Theta)\\
\langle + \mp\mp \rangle&=&\pm \frac{2m_t}{\sqrt{s}}v_R \sin\Theta ,
\end{eqnarray}
and $v_{L,R}$ and $a_{L,R}$ are
the standard vector and axial couplings of the
$\gamma$ and $Z$ to the electron and top quark currents. Numerical values
of $P$, $\bar P$ and $Q$, in units of $\Sigma_{unpol}$, at $\sqrt{s}$=
500 GeV and integrated over $\Theta$ are
\begin{equation}
\Sigma_{unpol}\;:\; P\;:\; \bar P\; :\;Q\;=\; 1\;:\;-0.18\;:\;0.18\;
:\;-0.63
\end{equation}
Thus, the spin-spin correlation term $Q$ in eq.(5) is expected to be
significant; it is found to be about four times
larger than the polarization function $P$. The ratio $Q/P$ depends weakly
on $\sqrt{s}$ in the range from 360 to about 1000 GeV,
so that for a $\Sigma_{unpol}$ variation in this energy range
by approximately
a factor of two to three, our analysis is not
critically dependent on $\sqrt{s}$. Throughout the paper
we have chosen $\sqrt{s}=$ 500 GeV.

The polarization degrees $k$ and $\bar k$ of the $t$ and $\bar t$ decay
amplitudes, summed over the $W$ helicity states,
are defined by the structure of the $Wtb$ vertex. If the spin 
quantization axis is collinear to the top momentum, the
$t\to W^+ b$ amplitude polarization density matrix in the rest frame of
the top has the form (\cite{KLY}, see details in the Appendix)
\begin{eqnarray}
\frac{1}{2} \left( \begin{array}{cc}
1+k\cos\theta & k\sin\theta e^{i\varphi} \\
k\sin\theta e^{-i\varphi} & 1-k\cos\theta
\end{array} \right)
\end{eqnarray}
The explicit expression for the polarization degree $k$ for the
$t\to W^+ b$ decay can be obtained in models
with the general effective lagrangian (1) by means of the eight helicity
amplitudes of the top decay defined in the Appendix. 
In the case $f_{1L}$=1, $f_{1R}$=0 we get
\begin{eqnarray}
k=\frac{(\frac{m_t}{m_W}+f_{2L})^2
-2(1+\frac{m_t}{m_W}f_{2L})^2-(1-2(\frac{m_t}{m_W})^2)\,f^2_{2R}}
{(\frac{m_t}{m_W}+f_{2L})^2
+2(1+\frac{m_t}{m_W}f_{2L})^2+(1+2(\frac{m_t}{m_W})^2)\,f^2_{2R}}
\end{eqnarray} 

The expressions (4), (5), (16) provide the basis for a
qualitative understanding of the results from exact matrix
element Monte Carlo calculations, when nonzero bottom quark
mass and finite top quark and $W$-boson widths are accounted for.

It follows from (5) that natural integrated angular observables are
the $b$-quark and the lepton forward-backward asymmetries, measured
in the rest frame of the top. It is straightforward
to show that these asymmetries have the form
\begin{equation}
A_{FB}=\frac{\sigma(\theta<90^{\circ})-\sigma(\theta>90^{\circ})}
            {\sigma(\theta<90^{\circ})+\sigma(\theta>90^{\circ})}=
\frac{k}{2}\,\frac{P}{\Sigma_{unpol}}
\end{equation}
For the $b$-quark, the polarization degree $k$ in the SM
equals 0.41 (see the Appendix) and the ratio $P/\Sigma_{unpol}$ (the
degree of longitudinal top quark polarization integrated over $\Theta$)
is equal to 0.18
\footnote
{The P dependence on the top production angle can be found in 
\cite{KRZ,AS}.}
at $\sqrt{s}=$ 500 GeV.
Hence, $A^b_{FB}$ in the infinitely small width approximation equals 
3.6\%, while for the lepton from $W$ decay, with $k$=1 in the SM,
$A^l_{FB}=$-9.0\%.

\subsection{Infinitely small $W$ width approximation in
the top quark anomalous decay}

The effective lagrangian terms of the $Wtb$ vertex can significantly
change the top quark two-body and three-body decay widths
if the anomalous couplings $f_{2L}$, $f_{2R}$ are sufficiently large.
Whether however finite $W$ width corrections substantially obscure
effects of anomalous couplings demands to investigate computations
done within approximation (3) and to reveal
its relation to exact Breit-Wigner
propagator calculations.
In order to quantify
this question we performed an explicit symbolic calculation of
the factorized branching ratios in formula (4). The result
for the two-body top decay width can be obtained
from the helicity amplitudes (30)-(37):
\begin{eqnarray}
\Gamma_2 (t\to W^+ b)=\frac{G_F\, m^3_t}{8\sqrt{2}\, \pi}
(1-r^2)^2 \ [1+2r^2+6\,f_{2L}r+(f^2_{2L}+f^2_{2R})(2+r^2)]
\end{eqnarray}
where $r=m_W/m_t$. The three-body top decay width, after
integration of the symbolic expression
over the Dalitz plot, is given by
\begin{eqnarray}
\Gamma_3 (t\to e^+ \nu_e b)=&\frac{G^2_F\, m^3_t\,m^2_W}{96\pi^3}
   \Bigl[ F_1\,
\frac{m_W}{\Gamma_W}(\pi
         -\arctan\frac{m^2_t\Gamma_W}{m_W(m^2_t-m^2_W-\Gamma^2_W)})
\\ \nonumber
 &+F_2
\,\log\frac{m^2_W(m^2_W+\Gamma^2_W)}{(m^2_t-m^2_W)^2+m^2_W\Gamma^2_W}
  + F_3 \Bigr]
\end{eqnarray}
where
\begin{eqnarray*}
F_1&=& 1-3r^4+2r^6+3r^2\gamma^2-6r^4\gamma^2 \\ 
  & & -6f_{2L}(2r^3-r^5-\gamma-2r\gamma^2+3r^3\gamma^2)\\
 & &+(f^2_{2L}+f^2_{2R})(2-3r^2+r^6+3\gamma^2-6r^4\gamma^2+r^2\gamma^4)\\
F_2&=&3r^4-3r^6+r^4\gamma^2-3f_{2L}(r-4r^3+3r^5-r^3\gamma^2)\\
  &  &+(f^2_{2L}+f^2_{2R})(-1+3r^2-2r^6+2r^4\gamma^2)\\
F_3&=&-3r\,f_{2L}(3-4r^2)+
               (f^2_{2L}+f^2_{2R})(-\frac{1}{3}+r^2+3r^4-r^2\gamma^2)
\end{eqnarray*}
and $\gamma=\Gamma_W/m_t$. If we set $f_{2L}=f_{2R}=0$ and use
the approximation $F_1=1-3r^4+2r^6$ and neglect $\Gamma_W/m_t$
power terms, we obtain by comparing (18) and (19)
the explicit narrow width factorization in the SM case:
\begin{eqnarray}
\lefteqn{\Gamma_3 (t\to e^+ \nu_e
b)=\frac{G^2_F\,m^3_t\,m^3_W}{96\pi^3}\frac{\pi}
{\Gamma_W}(1-3r^4+2r^6)=} \\ \nonumber
& & = \frac{G_F\,m^3_t}{8\sqrt{2}\pi}(1-r^2)^2 (1+2r^2)   
                    \,\frac{1}{\Gamma_W}\,
      \frac{G_F\,m^3_W}{6\sqrt{2}\pi}
=\Gamma_2 (t\to W^+b) \, Br(W^+\to e^+ \nu_e)
\end{eqnarray}
Since the $W$ branching factorisation (20) is in general not
valid, its violation by
$f_{2L,R} \cdot \Gamma_W/m_t$ and $f_{2L,R} \cdot m_W/m_t$ power terms 
is only weak provided modulus of $f_{2L}$ and $f_{2R}$
are around or less than 1.
More details about the precision of the factorization approximation
can be obtained from Fig.1, where the ratio
$\Gamma_3/(\Gamma_2\cdot Br(W^+\to e^+ \nu_e))$ as a
function of the anomalous
couplings $f_{2L}$ and $f_{2R}$ is shown. Clearly, 
the accuracy of $\Gamma_3$ within the infinitely small $W$
width approximation is convincing in the range considered
for $f_{2L}$ and $f_{2R}$;
deviations are expected to be of the
order 1\% or less. Thus, calculations done within the approximation (3)
imply small corrections which are less important than e.g.
interferences between the signal diagrams (see below). In general,
however, careful investigations are appropriate when anomalous
top quark decay calculations are carried out within the infinitely 
small $W$ width approximation.

\section{Tree-level results for $e^+ e^- \to t\bar t\to t l \bar \nu_l
\bar b$}

If precise measurements of top decay products are
envisaged, it is demanding to know the SM predictions with very
high accuracy. The program package CompHEP \cite{CompHEP}
which performs analytic calculations of the matrix element squared,
generates an optimized FORTRAN code and generates an event flow, 
overcomes the shortcomings due to infinitely small width and
zero fermion mass approximations. Furthermore, it allows to include all
diagrams of the irreducible background and their interferences.
In the case of the signal process, $e^+ e^- \to t \bar t \to t l \bar
\nu_l \bar b$,
only two diagrams and their interference exist. If the anomalous couplings
$f_{2L}$ and $f_{2R}$ are allowed to contribute to the lagrangian,
the corresponding Feynman rules implemented
in CompHEP can be found in the Appendix of the second ref. in 
\cite{boos1}.

\subsection{Forward-backward asymmetries}

It follows from eqs.(4) and (5) that observables of experimental interest
are the distributions in $\theta$, $\theta^*$ for the $b$-quark and 
lepton in the top rest frame, or in the $e^+ e^-$ center-of-mass system
(c.m.s.) for a more general discussion. From these distributions
the forward-backward asymmetry (17) can be easily calculated in
the SM and in models extended by anomalous couplings.

\unitlength 1cm
\begin{figure}[htb]
\begin{picture}(17,17)
\put(1,0){\epsfxsize=12cm
         \epsfysize=15 cm \leavevmode \epsfbox{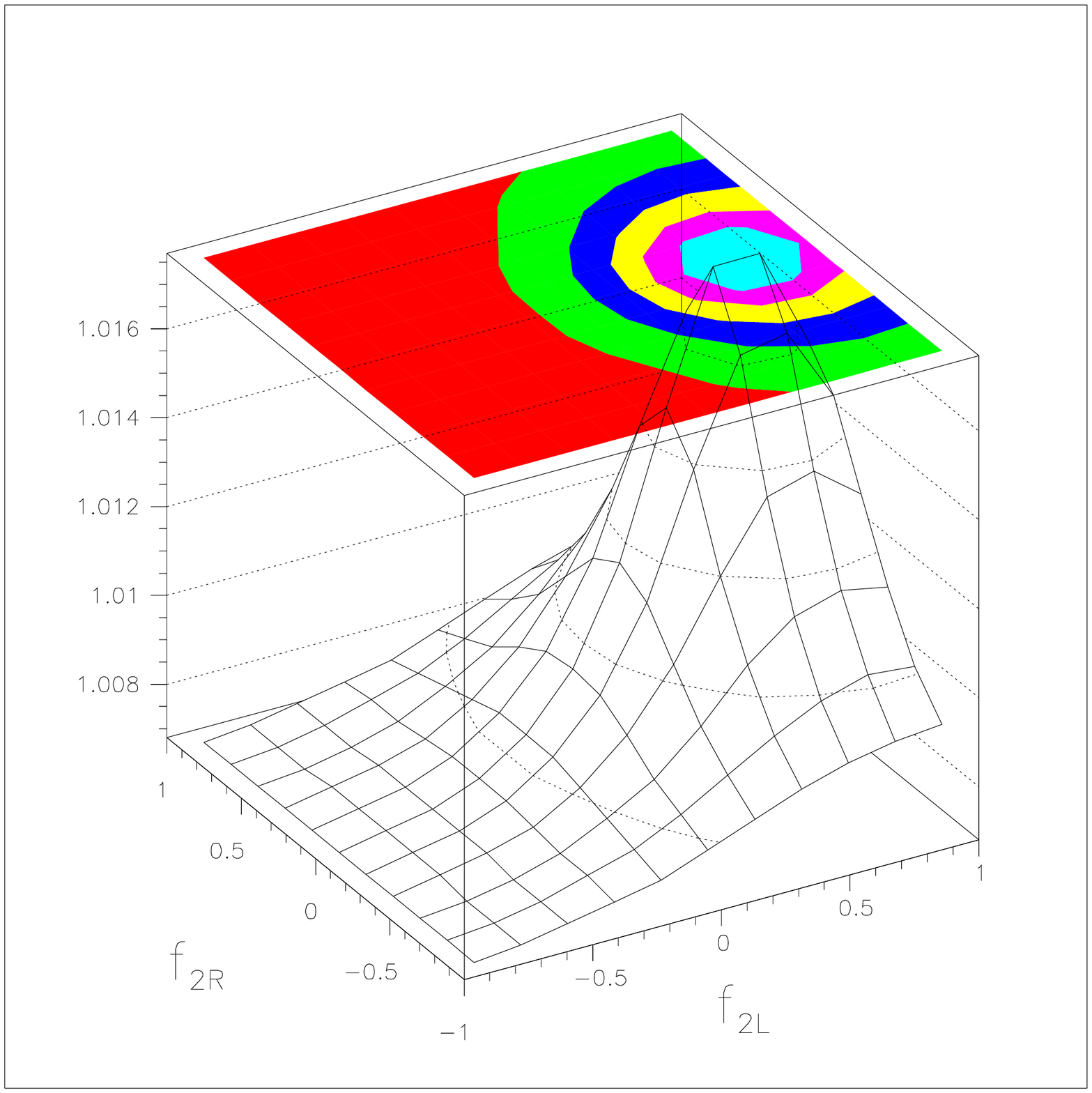}}
\end{picture}
\caption{
The ratio $\Gamma_3 (t \to e^+ \nu_e b)/
(\Gamma_2 (t \to W^+ b)\, Br(W^+ \to e^+ \nu_e))$ versus $f_{2L}$ and
$f_{2R}$, with equidistant isocontours in the $f_{2L}, f_{2R}$
projection.}
\end{figure}

In a first step we compare precise calculations with results
obtained from the infinitely small width approximation (3),
within the SM ($f_{2L}=f_{2R}=$0). Numerical values for
the $b$-quark and lepton asymmetries, $A^b_{FB}$ and $A^l_{FB}$,
calculated by means of CompHEP, are shown in Table 1,
in the top rest frame as well as in the $e^+ e^-$ c.m.s.
If compared with the top rest frame asymmetries obtained within
the narrow width approximation of sect.3.1, one notices
a 15\% difference for the $b$-quark,
whereas for the lepton the difference is negligible. Thus,
already this example demonstrates the importance of precise
calculations which include
interference terms, finite width and non-zero mass contributions.

The SM forward-backward $b$-quark asymmetries (Table 1)
are significantly larger in the $e^+ e^-$ c.m.s. than in the
top rest frame, while for the lepton such differences are less
evident.
This observation can be understood by recalling that the $t$ ($\bar t$) is
produced mainly in the $e^-$ ($e^+$) direction with left (right)
helicity and, in the top decay, the lepton ($b$ quark) is emitted
preferrably in the direction of (in the opposite direction to) the top
spin.

\begin{table}[t]
\begin{center}
\begin{tabular}{|c|c|c|c|c|}
\hline
 & $f_{2R}$& $f_{2L}$&$A_{FB}$, $e^+ e^-$ c.m.s. & $A_{FB}$, top frame \\
\hline 
\multicolumn{5}{|c|}{}                             \\ \hline
\multicolumn{1}{|c|}{} & \multicolumn {4}{|c|}{unpolarized $e^+ e^- \to
                                  t \mu \bar \nu_{\mu} \bar b$}
                        \\ \hline
$\bar b$ & 0.0 & 0.0 & 0.279 &  0.030      \\
$\bar b$ & 0.0 &-0.2 & 0.243 &  0.010      \\
$\bar b$ & 0.0 &-0.4 & 0.218 & -0.004      \\
$\bar b$ & 0.0 &-0.6 & 0.197 & -0.020      \\ 
$\bar b$ & 0.0 &-1.0 & 0.169 & -0.039      \\  \hline
$\bar b$ &-0.6 & 0.0 & 0.301 &  0.041      \\
$\bar b$ &-1.0 & 0.0 & 0.315 &  0.045      \\ \hline 
$\mu   $ & 0.0 & 0.0 & 0.079 & -0.091     \\
$\mu   $ & 0.0 &-0.6 & 0.085 & -0.084     \\  \hline
\multicolumn{1}{|c|}{} & \multicolumn {4}{|c|}{polarized $e^-_L e^+ \to
                                  t \mu \bar \nu_{\mu} \bar b$}
                         \\ \hline
$\bar b$ & 0.0 & 0.0 & 0.354 &  0.100       \\
$\bar b$ & 0.0 &-0.2 & 0.265 &  0.034      \\
$\bar b$ & 0.0 &-0.4 & 0.200 & -0.011      \\
$\bar b$ & 0.0 &-0.6 & 0.152 & -0.047       \\
$\bar b$ & 0.0 &-1.0 & 0.087 & -0.095     \\    \hline
$\mu   $ & 0.0 & 0.0 & 0.145 & -0.262     \\ 
$\mu   $ & 0.0 &-0.6 & 0.104 & -0.233     \\ \hline

\end{tabular}
\end{center}
\caption{Forward-backward asymmetries for the
$b$-quark and lepton in the reaction $e^+ e^- \to t l^- \bar \nu_l
\bar b$ at $\sqrt{s}=$500 GeV, for the standard and
anomalous effective $Wtb$ vertices,
calculated in the $e^+ e^-$ center of mass frame and in the
top rest frame.}
\end{table}

\begin{figure}[htb]
\begin{picture}(17,17)
\put(1,0){\epsfxsize=12cm
         \epsfysize=15 cm \leavevmode \epsfbox{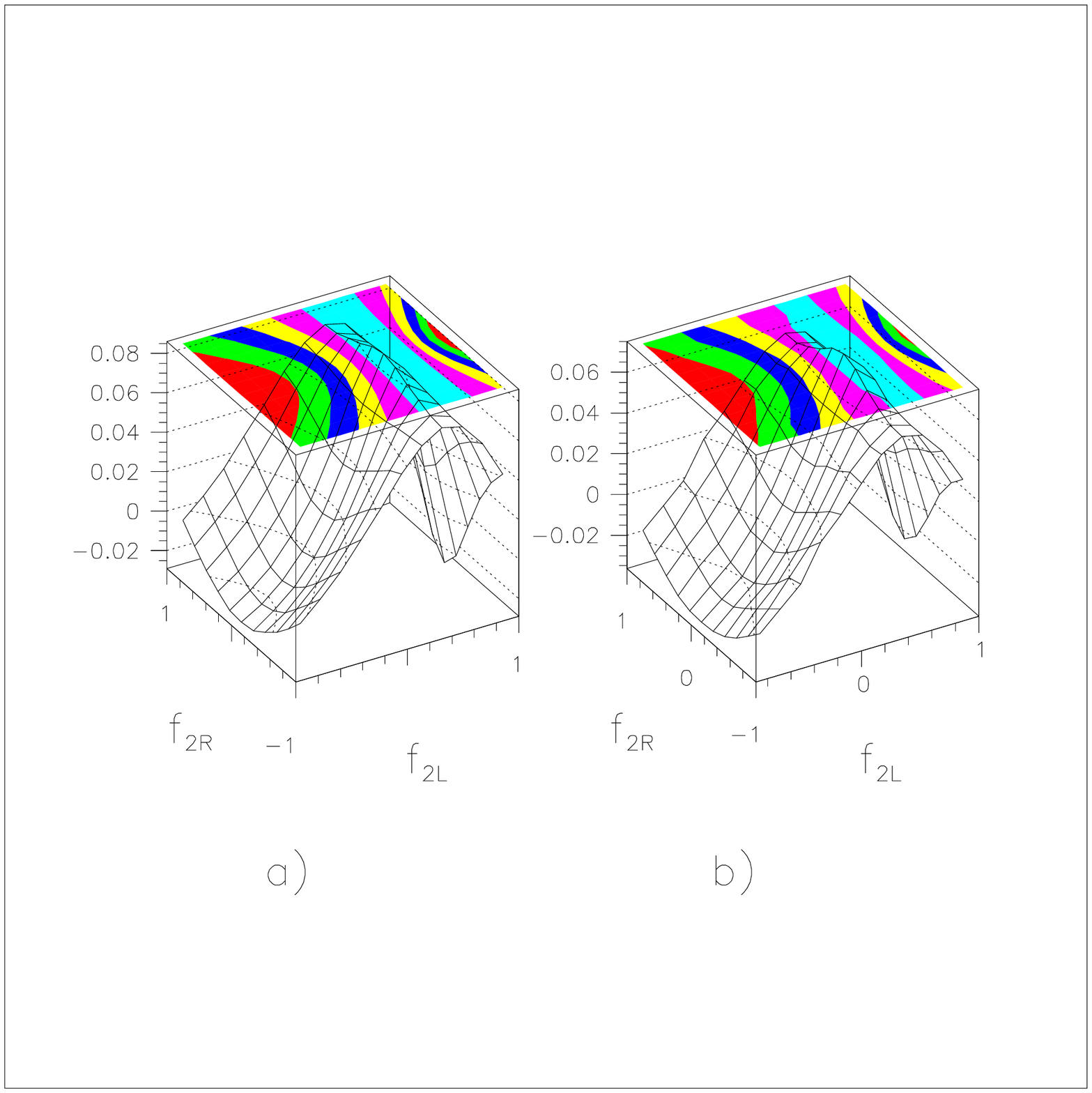}}
\end{picture}
\caption{ Forward-backward $b$-quark asymmetry in the
top rest frame
a) calculated within the infinitely small width
approximation and b) precisely calculated using CompHEP, for
the reaction $e^+ e^- \to t\bar t\to t l^- \bar \nu_l \bar b$
as a function of $f_{2L}$ and $f_{2R}$. $A^b_{FB}$-equidistant isocontours
are also shown in the $f_{2L}, f_{2R}$ projection.}
\end{figure}

It is also worth to mention that irreducible background,
which might remain after any $t \bar t$ selection procedure,
should be carefully accounted for.
CompHEP calculation shows that
if the electron being the lepton in the final state
(with 18 contributing diagrams in total) forward
electrons from $t$-channel photon exchange alter significantly 
$A^e_{FB}$ compared to only signal diagrams calculations.

When we allow for anomalous $Wtb$ couplings, the asymmetry
$A^b_{FB}$, measured in the top rest frame, is shown in Fig. 2 in the
narrow width approximation  and for exact calculations.
The qualitative behaviour of both asymmetries as a function
of $f_{2L}$ and $f_{2R}$ is very similar;
only close inspections reveal significant differences.
Furthermore, $A^b_{FB}$
depends stronger on $f_{2L}$ than on $f_{2R}$, as can be better seen
in Fig.3a, where also two standard exclusion contour plots
are shown for 100 fb$^{-1}$ and 500 fb$^{-1}$ integrated luminosities.
This greater
sensitivity is directly connected to the stronger influence of the linear
$f_{2L}$ term in (16) than that of the quadratic $f_{2R}$ term, which
in turn is an inherent property of the helicity amplitudes of anomalous top
quark decays, as outlined in the Appendix.

The impact of the anomalous couplings to the lepton forward-backward
asymmetry $A^l_{FB}$ is less important both in the $e^+ e^-$ c.m.s. and
the top rest frame (see Fig.3b), and being only indirect due
to the presence of the standard left current $W$ boson decay.
However, in contrast to the two-fold ambiguity of the b-quark
asymmetry (Fig. 3a), $A^l_{FB}$ is unique in the sense that
for a given $f_{2R}$ value only one $f_{2L}$ range (shaded) exists,
in which no distinction from the SM (within $2\sigma$) is possible.

Table 1 contains some numerical examples of
$A^{b/l}_{FB}$ for several $f_{2L}$
and $f_{2R}$ values at $\sqrt{s}=$500 GeV,
measured in both reference frames discussed so far. Clearly, the
largest forward-backward asymmetry is obtained for the
$b$ quark if measured in the $e^+ e^-$ c.m.s. One
should however remember that $e^+ e^-$ c.m.s. asymmetries are a superposition of
production and decay asymmetries, while asymmetries
reconstructed in the top rest frame can be considered
as a 'pure' effect.

Left electron beam polarization not only increases the $t\bar t$
production rate by a
factor of about three ($\sigma_{tot}=$176 fb
for 100\% left polarized electrons at
$\sqrt{s}=$500 GeV) but also enhances $A^{b/l}_{FB}$ by a factor of 2-3 
in the top rest frame. At the same time, the $f_{2L}$
sensitivity of $A^{b}_{FB}$ increases most significantly
if measured in the $e^+ e^-$ c.m.s.
(see the examples in Table 1). 

Besides the study of the $\theta$ and
$\theta^*$ decay angular distributions, more
sophisticated angular
observables were proposed to study parity violating effects:
(1) the angle between the lepton momentum in the $W$ rest
frame and the momentum of the top in the $e^+ e^-$ c.m.s. \cite{LY} and
(2) the angle between the top production plane and the
production plane of the $b$-quark (or the lepton)
in the $e^+ e^-$ c.m.s.:
\begin{equation}
\cos \theta_{tb}= \frac{([\bf{ k t}] \cdot [\bf{ k b}])}
                      {|[\bf{ k t}]||[\bf{ k b}]|} \hspace{0.5cm},
\end{equation}
where $\bf{k}$ is a unit length vector in the $e^-$ direction.
A similar variable was proposed in \cite{KPR} to measure the
transverse quark polarization.

In both angular distributions very large asymmetries (up to 95\%) exist.
However, their sensitivity to the anomalous $Wtb$ couplings 
$f_{2L}$ and $f_{2R}$
is very small and is, in good approximation, independent
on the electron polarization.

\subsection{Energy spectrum asymmetry}
Besides angular distributions, energy spectra of the top decay
products may also possess high sensitivity to anomalous
couplings. In this section we study the asymmetry 
of the lepton energy spectrum defined in the
top rest frame using the dimensionless variable
$x_{\mu}=2\, E_{\mu}/m_{top}$:
\begin{eqnarray}
A^{\mu}_E = \frac{ \sigma(x_{\mu}<0.5)-\sigma(x_{\mu}>0.5)}
                 { \sigma(x_{\mu}<0.5)+\sigma(x_{\mu}>0.5)} 
\end{eqnarray}
Fig.3c shows the results for $A^{\mu}_E$ from the process
$e^+ e^- \to t \mu \bar \nu_{\mu} \bar b$ as a function
of $f_{2L}$ and $f_{2R}$. As can be seen, $A^{\mu}_E$ is significantly more
sensitive to $f_{2L}$ than its forward-backward asymmetry
$A^{\mu}_{FB}$, and this result is independent whether $A^{\mu}_E$
is measured in the top rest frame or in the $e^+ e^-$ c.m.s.
Alike to $A^{b}_{FB}$, the sensitivity to $f_{2R}$ is somewhat
less pronounced than to $f_{2L}$ and the ambiguity exists also.

The $b$-quark energy spectrum in the top rest frame has a resonance peak
at $x_b=1-(m_W/m_t)^2$, resulting to an energy asymmetry insensitive to
anomalous couplings.

If the neutrino is used as the analyser (by means of the missing
energy technique), its energy asymmetry is slightly
less sensitive to $f_{2L}$ and $f_{2R}$
than the lepton energy asymmetry. Whether the neutrino
is at all usable for precise measurements, requires however detailed
experimental studies including full event simulation
and reconstruction.

As clearly visible from Figs.3a-c, only the combination of
forward-backward and energy asymmetry measurements results to
an allowed region (Fig.4, the sum of the grey and dark areas) much
smaller than that for each measurement alone and ensures a
significant improvement of the sensitivity on
anomalous couplings, with limits for $f_{2L}$ and $f_{2R}$
sensible luminosity dependent.

\subsection{Spin-spin asymmetries}

The spin correlations and the spin-spin asymmetries, which
are related to each other, are
in general double differential distributions where one of the
variables is integrated over a certain kinematical region. As
already mentioned in sect. 3, the spin correlation term $k\bar k Q$ in
(5) is comparable to the polarization term $kP$ and therefore spin-spin 
correlations, although suppressed by the additional power of $k$, are
expected to be not small.  
For instance, the forward-backward asymmetry of the $b$-quark
measured in the top rest frame, under the condition that the $\bar
b$-quark is observed only in the forward hemisphere in the $\bar t$ rest
frame, can be derived from (5) as
\begin{equation}
A^b_{s}=\frac{k}{2}\frac{P}{\Sigma_{unpol}}(1+\bar k \, \frac{Q}{2P}) \hspace{0.5cm}.
\end{equation}
For simplicity, the term $\bar k \bar P/\Sigma \ll$1 is omitted here.
At e.g. $\sqrt{s}=$500 GeV, we find $A^b_{s}=$0.062 within the SM, 
which is two times larger than $A^b_{FB}=$0.036, when no restriction
is imposed on the $\bar t$ side. 

If anomalous couplings are allowed to contribute,
the dependence of the spin-spin asymmetry, $A^b_{s}$, on
$f_{2L}, f_{2R}$ is shown in Fig.5,
within the narrow width approximation. 
Clearly, improved constraints
on $f_{2L}, f_{2R}$ can be obtained from $A^b_{s}$ compared to the
unrestricted forward-backward asymmetries as discussed in sect. 4.1.
However, since $A^b_{s}$ is calculated in the infinitely small width
approximation 
\footnote{For reasons of unsufficient computer memory,
CompHEP $2\to 6$ calculations are only possible for the SM
$Wtb$ vertex, with the result $A^b_{s}=$0.049 at $\sqrt{s}=$500 GeV.},
the reliability of the results needs more careful investigation.

\begin{figure}[h]
\begin{picture}(17,17)
\put(1.2,0){\epsfxsize=13cm
         \epsfysize=17 cm \leavevmode \epsfbox{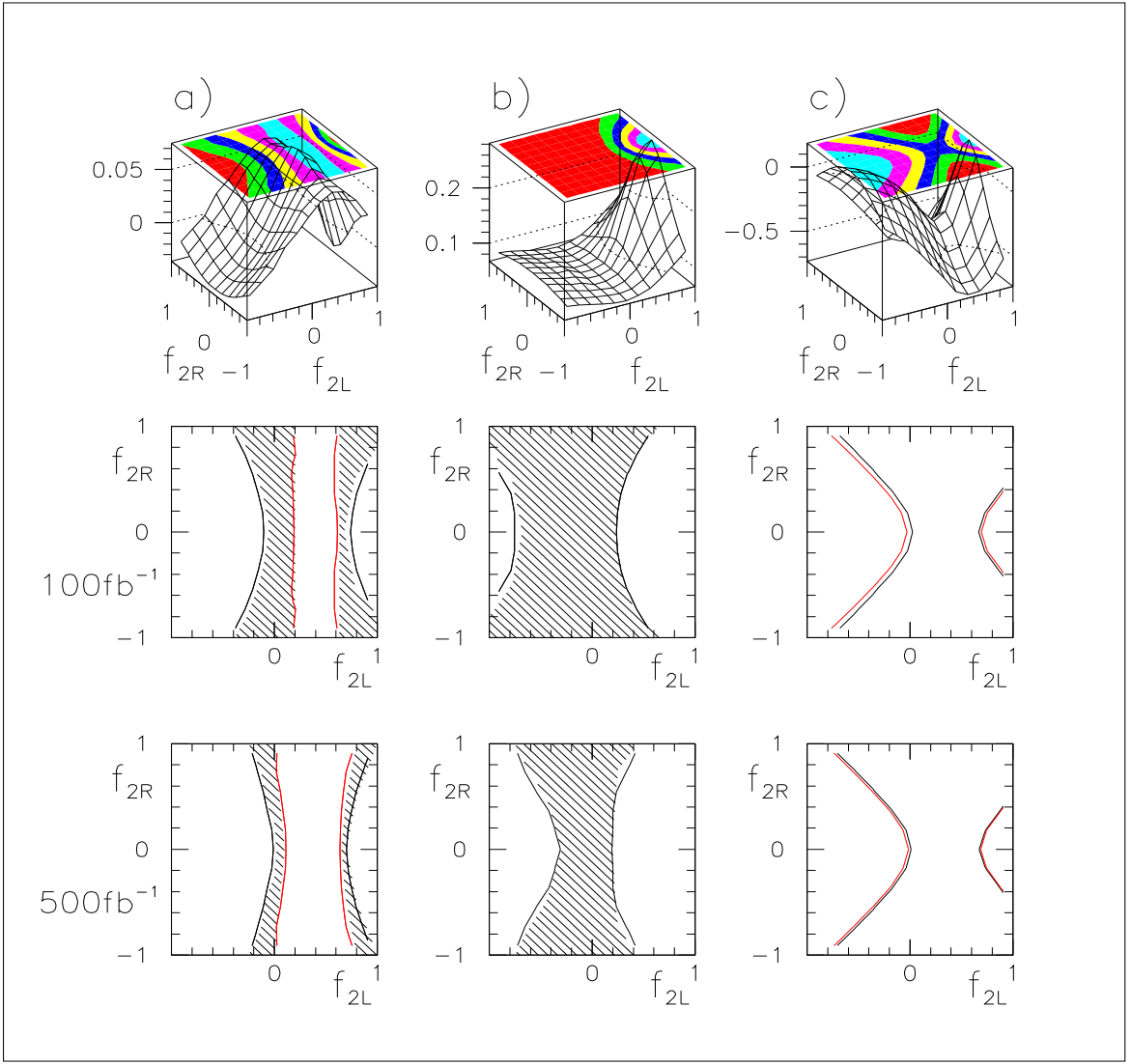}}    
\end{picture}
\caption{
a) Forward-backward $b$-quark asymmetry in the top rest
frame, b) forward-backward lepton asymmetry in
the $e^+ e^-$ c.m.s. and c) energy asymmetry for the
lepton in the top rest frame, as functions of $f_{2L}$ and $f_{2R}$,
for the reaction $e^+ e^- \to t\bar t\to t l^- \bar \nu_l \bar b$ at
$\sqrt{s}=$500 GeV. Also shown are the 2$\sigma$ limits on the
anomalous couplings  of each observable (shaded),
for integrated luminosities of 100 fb$^{-1}$
and 500 fb$^{-1}$, and $A^b_{FB} (A^l_{FB}, A^l_{E})$-equidistant
isocontours in the $f_{2L}, f_{2R}$ projection.}
\end{figure}

\begin{figure}[h]
\begin{picture}(17,17)
\put(1,0){\epsfxsize=12cm
         \epsfysize=15 cm \leavevmode \epsfbox{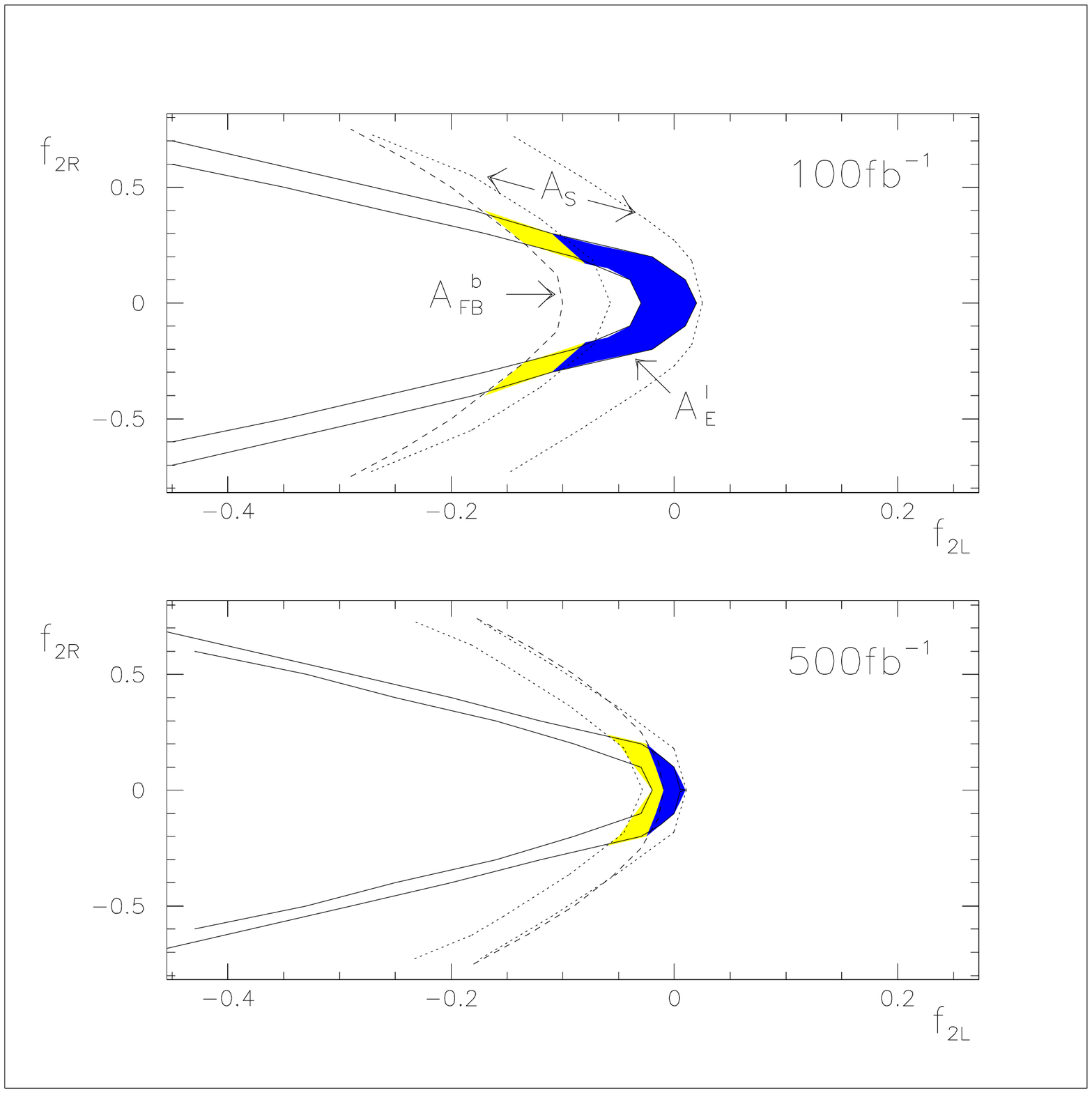}}
\end{picture}
\caption{
Combined 2$\sigma$ limits on the anomalous couplings
$f_{2L}$ and $f_{2R}$ of the reaction $e^+ e^- \to t\bar t\to t l^- \bar
\nu_l \bar b$ $ ((l^- \bar \nu_l \bar b)(W^+ b))$ at
$\sqrt{s}=$500 GeV, for 100 fb$^{-1}$ and 500 fb$^{-1}$
integrated luminosities. For the meaning of the
gray and dark areas we refer to the text.
}
\end{figure}

We expect some further enhancement of parity-violating effects by using
polarized beams. If e.g. 100\%
left polarized electrons collide with unpolarized positrons, forward
$t$ (backward $\bar t$) quarks (w.r.t. the $e^-$ direction)
are mainly produced in the
helicity configuration L (R), while backscattered $t$ (forward
$\bar t$) are produced in the helicity configuration R (L)
(their production angular behaviour can be found in \cite{spincorr}).
As a consequence, the $t$ and $\bar t$ decay products in
the reaction
\begin{eqnarray}
e^+ e^- \to t \bar t \to (e^- \nu_e \bar b) (u \bar d b)
\nonumber
\end{eqnarray}
are expected to be strongly correlated, in so far as
the $e^-$ and the $\bar d$ are produced mainly in the top spin direction,
while the u- and b-quarks prefer production in the opposite
direction \cite{topdecay}.
Using CompHEP to calculate
the exact $2 \to 6$ SM amplitudes, we obtain the electron
decay angular distribution in the top rest frame
under the condition that the $\bar d$ decay angle
in the antitop rest frame is less than 90$^{\circ}$,
for the top produced either forward (Fig. 6a)
or backward (Fig. 6b) in the $e^+ e^-$ c.m.s.
The spin-spin asymmetries in these two cases are -0.258 and
-0.096, respectively, demonstrating strong sensitivity to
parity-violating effects when beam polarization is available. 
QCD corretions to the
spin correlations in $t\bar t$ production are expected
to be in general small \cite{BFU}, but their inclusion is recommended in
searches for nonstandard interactions.

\section{Conclusions}

The total rate of the reaction 
$e^+ e^- \to t \bar t \to $ 6  fermions at NLC energies
is negligibly affected by the anomalous lagrangian terms (1).
Hence, it is straightforward to use single and double differential
distributions of the top/antitop decay products
to eventually observe effects
due to anomalous $Wtb$ operators, and as larger their sensitivity
as stronger limits on anomalous couplings can be imposed.

In this paper we investigate forward-backward
asymmetries for the b-quark and the lepton in the top rest frame
or in the $e^+ e^-$ c.m.s., the energy asymmetry
 of the lepton in the top rest
frame and the spin-spin asymmetry in the $t$/$\bar t$ decay.
Precise tree-level Monte Carlo calculations
for the signal diagrams and their interference were performed and
compared in the case of forward-backward asymmetries
with the symbolic expressions obtained in the infinitely
small width and zero fermion mass approximation.
We realized that in general
careful investigations are appropriate when such
approximations are intended to be used in analyses of multiparticle
final state topologies.

\begin{figure}[h]
\begin{picture}(17,17)
\put(1,0){\epsfxsize=12cm
         \epsfysize=15 cm \leavevmode \epsfbox{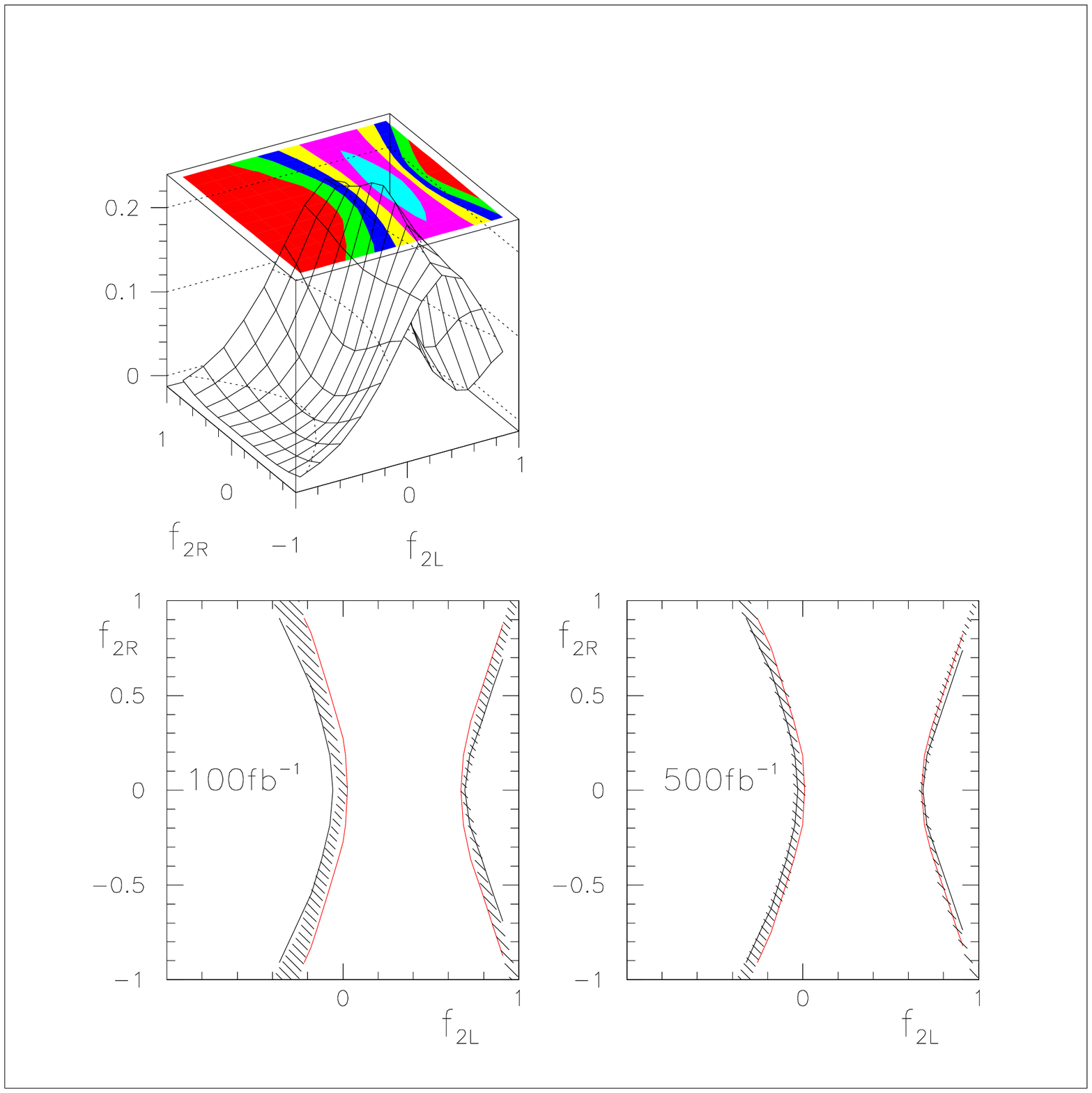}}
\end{picture}
\caption{
Spin-spin asymmetry for the $b$-quark in the top rest frame as
a function of $f_{2L}$ and $f_{2R}$ for the reaction
$e^+ e^- \to t \bar t \to (l^- \nu_l \bar b) (u \bar d b)$
at $\sqrt{s}=$500 GeV. Also shown are the 2$\sigma$ limits
on the anomalous couplings (shaded) for 100 fb$^{-1}$ and 500 fb$^{-1}$
integrated luminosity and $A^b_{S}$-equidistant isocontours in the
$f_{2L}, f_{2R}$ projection.} 
\end{figure}

\begin{figure}[h]
\begin{picture}(17,17)
\put(-1.5,6){\epsfxsize=17cm
         \epsfysize=10 cm \leavevmode \epsfbox{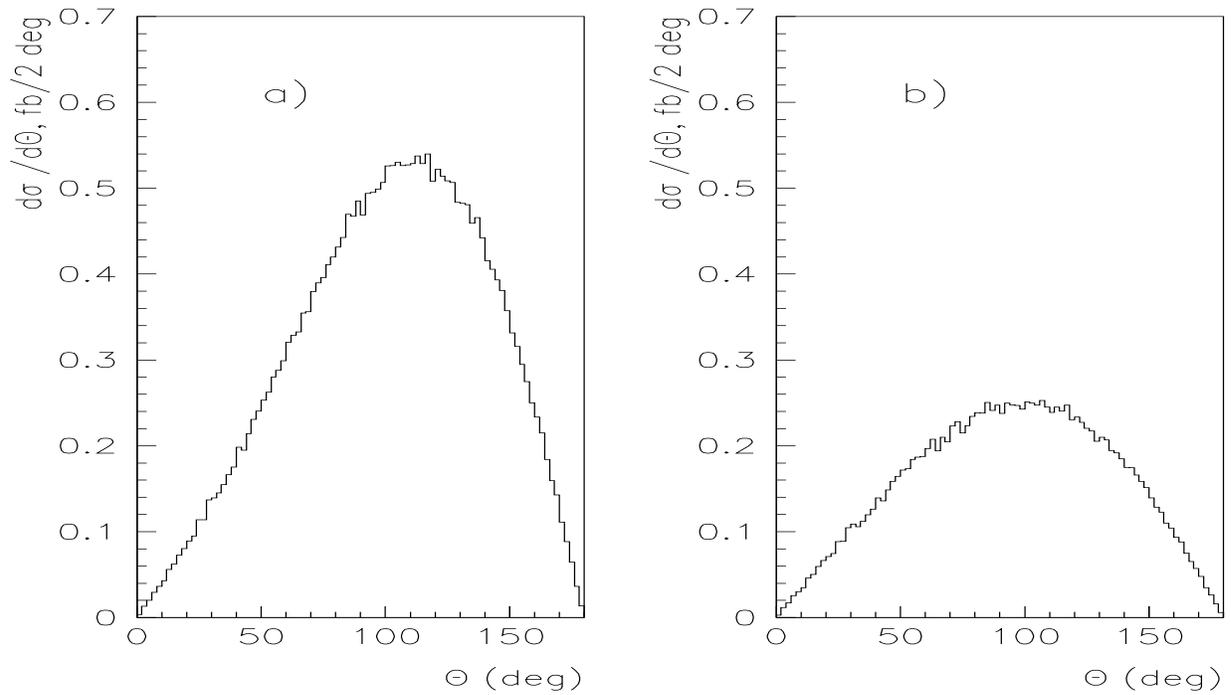}}
\end{picture}
\vspace {-5cm}
\caption{ 
Angular distribution between the muon and the top in
the top rest frame
under the conditions that the angle
between the $\bar d$ and the $\bar t$ in the $\bar t$
rest frame is less than 90$^{\circ}$ and a) the top is produced
in the forward hemisphere or b) the top is produced in the backward
hemisphere, for the reaction
$e^+_L e^- \to \mu^- \bar \nu_{\mu} b u \bar d  \bar b$ with
longitudinally polarized electrons at $\sqrt{s}=$500 GeV. 
}
\end{figure}

Concerning the sensitivity of the observables considered in this study
we found that (a) $A^b_{FB}$, $A^l_{FB}$ and $A^l_{E}$ have
stronger sensitivity to $f_{2L}$ than to $f_{2R}$, as seen in Fig.3
\footnote{
For $f_{1R}=$0, the
helicity amplitudes (30)-(37) have linear
and quadratic terms in $f_{2L}$ and only quadratic terms in $f_{2R}$.};
(b) the sensitivity of the forward-backward $b$-quark asymmetry
(Fig.3a) is larger than the sensitivity of the lepton
forward-backward asymmetry (Fig.3b),
which is somewhat degraded due to the subsequent $W$ decay; (c) it is
important to note that $A^l_{FB}$ resolves the ambiguity observed in
$A^b_{FB}$ and $A^l_{E}$; (d) the lepton energy asymmetry has the
largest sensitivity on $f_{2L}$ and $f_{2R}$ (Fig.3c).
In summary, it turns out that particle orientations
seem to be less sensitive to anomalous $Wtb$ operators than
particle energies.

As indicated by the 2$\sigma$ exclusion contour plots
in Figs. 3a-c, no satisfactory
restriction on $f_{2L}$ and $f_{2R}$ has been obtained for
each variable alone. But their combined annulus (Fig. 4) allows
significant improvements of the sensitivity on anomalous couplings.
If in addition the spin-spin asymmetry of Fig. 5, although calculated
within the narrow width approximation, is included,
further restrictions on anomalous $Wtb$ operators are
possible for 100 fb$^{-1}$ (dark area in Fig.4), while
for 500 fb$^{-1}$ no improvements are observed.
Thus, for
the high luminosity option of the TESLA linear collider \cite{B}
the bounds on the anomalous couplings $f_{2L}$ and $f_{2R}$,
within those no distinction from the SM is possible,
are [-0.025, 0] for $f_{2L}$ and $\pm0.20$ for $f_{2R}$.
These rather promising results demonstrate the reliability of
the top pair production process in $e^+ e^-$ collisions
to probe the $Wtb$ vertex.

It is interesting to compare these limits with the
expectations from single top production processes at LHC \cite{boos2}.
The LHC limitations, being 2-3 times better than the possible
restrictions from the upgraded Tevatron, are
comparable to the $e^+ e^-$ LC estimates
provided the LHC systematic uncertainties are controlled at a level
better than about 10\%.
The advantage of the LHC to measure the single top production
rates in the $Wb\bar b$/$Wb\bar b+jet$ channels \cite{boos2} is
however somewhat degraded by relatively large uncertainties
in the absolute normalization of the cross sections and the
presence of reducible background not easy to control.
In the clean environment of $e^+ e^-$ collisions, the selection
of $t\bar t$ events is thought to be very reliable and further
improvements in probing the $Wtb$ vertex can be
expected if additional sensitive observables are included in
the analysis and electron beam polarization is used.
Whether however the superior sensitivity to the anomalous couplings
$f_{2L}$ and $f_{2R}$ of a linear collider in the $\gamma e$ mode
at high energies ($\sqrt{s_{e\gamma}}\geq$1 TeV) \cite{boos1}
could be achieved or even superseded,
remains open for future studies.

\begin{center}
{\bf Acknowledgments}
\end{center}
E.B. and M.D. are grateful to DESY-Zeuthen for hospitality.
M.D. thanks very much H.S.Song for useful discussions.
The work of E.B. and M.D. was partially supported by the RFBR-DFG
grant 99-02-04011, the CERN-INTAS grant and the KCFE grant (SPb).

\section*{Appendix}

The helicity amplitudes for the decay $t \to W^+ b$ in models with
the general interaction lagrangian (1) can be found in \cite{KLY}. Our
calculation follows the formalism of \cite{HZ}, where the chiral
representation for the gamma matrices is used. The four component
spinors can be split into two component helicity eigenstates
$\chi_{\lambda}(p)$
\begin{eqnarray}
u(p,\lambda)_{\pm}=\omega_{\pm \, \lambda}(p)\chi_{\lambda}(p) \\
\nonumber
v(p,\lambda)_{\pm}=\pm\lambda \omega_{\mp \,
\lambda}(p)\chi_{-\lambda}(p) ,
\end{eqnarray}
where
\begin{eqnarray}
\omega_{\pm}(p)=\sqrt{E \pm p} \; . \nonumber
\end{eqnarray}
In the rest frame of the top, the helicity eigenstates
of the $b$-quark can be written in the form
\begin{eqnarray}
\chi_{+}(p_b)=  \left( \begin{array}{c} {\tt sin} \frac{\theta}{2}\\
-{\tt cos} \frac{\theta}{2} e^{i\varphi} \end{array} \right),
\hspace{10mm}
\chi_{-}(p_b)=  \left( \begin{array}{c} {\tt cos}
\frac{\theta}{2}e^{-i\varphi}\\
{\tt sin} \frac{\theta}{2} \end{array} \right) 
\end{eqnarray}
with the following component representation of $W$ and $b$ momenta
in the spherical coordinate system
\begin{eqnarray}
p_W &=&\{E_W, |{\bf p}_W|{\tt sin}\theta{\tt cos}\varphi, |{\bf
p}_W| {\tt
sin}\theta{\tt sin} \varphi,
           |{\bf p}_W|{\tt cos}\theta \} \\
p_b &=&|{\bf p}_b| \{1, {-\tt sin}\theta{\tt cos}\varphi, -{\tt 
sin}\theta{\tt sin} \varphi,
           -{\tt cos}\theta \}
\end{eqnarray}
The polarization vectors of the $W$ boson can be taken in the form
\begin{eqnarray}
\epsilon_{+}&= &\frac{1}{\sqrt{2}}
\{ 0, -{\tt cos}\theta{\tt cos}\varphi+i{\tt sin}\varphi,
      -{\tt cos}\theta{\tt sin}\varphi-i{\tt sin}\varphi,
       {\tt sin}\theta \} \\ \nonumber
\epsilon_{-}&= &\frac{1}{\sqrt{2}}
\{ 0,  {\tt cos}\theta{\tt cos}\varphi+i{\tt sin}\varphi,
       {\tt cos}\theta{\tt sin}\varphi-i{\tt cos}\varphi,
      -{\tt sin}\theta \} \\ \nonumber
\epsilon_{0}&= &\frac{E_W}{m_W} \{ \frac{|{\bf p}_W|}{E_W},
       {\tt sin}\theta {\tt cos}\varphi,
       {\tt sin}\theta {\tt sin}\varphi,
       {\tt cos}\theta \} 
\end{eqnarray}
In the symbolic calculations we always neglect the $b$-quark mass. The
eight helicity amplitudes $\sqrt{2 E_b m_t} \, \langle h_t, \, h_W, \,
h_b\rangle$
corresponding to the matrix element of the top decay
\begin{eqnarray}
&\frac{g}{\sqrt{2}}\bar u(p_b)\Gamma_{\mu} u(p_t) \varepsilon^*_{\mu}(p_W)
  \nonumber
\end{eqnarray}
with
\begin{eqnarray}
\Gamma_{\mu}&=&f_{1L} \gamma_{\mu} (1-\gamma_5)
         +f_{1R} \gamma_{\mu} (1+\gamma_5) \\ \nonumber
     && +\frac{f_{2L}}{2m_W}(\hat p_W \gamma_{\mu}-\gamma_{\mu} \hat p_W)
         (1+\gamma_5)
         +\frac{f_{2R}}{2m_W}(\hat p_W \gamma_{\mu}-\gamma_{\mu} \hat p_W)
         (1-\gamma_5) \nonumber
\end{eqnarray}
can be calculated in the rest frame of the top using (24)-(28):
\begin{eqnarray}
\langle -,0,- \rangle &=&(\frac{m_t}{m_W}f_{1L}+f_{2L}){\tt
sin}\frac{\theta}{2}
\\
\langle -,-,- \rangle &=&\sqrt{2}(f_{1L}+\frac{m_t}{m_W}f_{2L}){\tt
cos}\frac{\theta}{2} \\
\langle +,0,- \rangle &=&(\frac{m_t}{m_W}f_{1L}+f_{2L}){\tt
cos}\frac{\theta}{2}
e^{i\varphi} \\ 
\langle +,-,- \rangle &=&-\sqrt{2}(f_{1L}+\frac{m_t}{m_W}f_{2L}){\tt
sin}\frac{\theta}{2} e^{i\varphi} \\ 
\langle +,0,+ \rangle &=&-(\frac{m_t}{m_W}f_{1R}+f_{2R}){\tt
sin}\frac{\theta}{2} \\
\langle +,+,+ \rangle &=&\sqrt{2}(f_{1R}+\frac{m_t}{m_W}f_{2R}){\tt
cos}\frac{\theta}{2} \\
\langle -,0,+ \rangle &=&(\frac{m_t}{m_W}f_{1R}+f_{2R}){\tt
cos}\frac{\theta}{2}
e^{-i\varphi} \\ 
\langle -,+,+ \rangle &=&\sqrt{2}(f_{1R}+\frac{m_t}{m_W}f_{2R}){\tt
sin}\frac{\theta}{2} e^{-i\varphi} 
\end{eqnarray} 
The sum of the eight helicity amplitudes squared gives the total decay
width of the top (18) for the general interaction lagrangian (1).
If $f_{1R}$=0, the width $\Gamma(t \to W^+ b)$ contains a linear
term in $f_{2L}$ and quadratic terms in both $f_{2R}$ and $f_{2L}$.
The eight helicity amplitudes of the antitop decay, $\bar t \to W^-
\bar b$, can be obtained from (30)-(37) by the replacements
$f_{1L}\leftrightarrow f_{1R}$ and $f_{2L}\leftrightarrow f_{2R}$ 
(only real $f$ are considered). 
In the Standard Model (SM) with $f_{1L}$=1, $f_{1R}=f_{2L}=f_{2R}$=0
only four nonvanishing helicity amplitudes remain \cite{BCK}.

If $(\theta,\varphi)$ are the polar and azimutal angles of the
$b$-quark with respect to the top momentum, the helicity amplitudes of the
top with spin up and spin down, $a_1$ and $a_2$, allow to define
the $t\to W^+ b$ {\it amplitude} polarization density matrix \cite{KLY}.
This matrix is different from the
polarization density matrix defined by the individual top spin function.
The squared sum of (32)-(35) gives the probability of spin up top
decay, while the probability of spin down top decay is given by the
squared sum of (30)-(31) and (36)-(37). 
The polarization density matrix can be defined for the
($a_1$,$a_2$) spin function
\begin{equation}
 \rho=\, \left( \begin{array} {cc}
|a_1|^2 & a_1\, a^*_2\\
a^*_1\, a_2 & |a_2|^2   \end{array}     \right) \; ,
\end{equation}
where the normalised SM components derived from (30)-(33) have the form
\begin{eqnarray*} 
a^2_1&=&\frac{M^2_W}{m^2_t+2m^2_W}\,(
\frac{m^2_t}{m^2_w}{\tt cos}^2\frac{\theta}{2}+2{\tt sin}^2
\frac{\theta}{2}) \\
a^2_2&=&\frac{M^2_W}{m^2_t+2m^2_W}\,(
\frac{m^2_t}{m^2_w}{\tt sin}^2\frac{\theta}{2}+2{\tt 
cos}^2\frac{\theta}{2}) \\
a_1\,a^*_2=a^*_1\, a_2&=&\frac{m^2_t}{m^2_W}{\tt sin}^2\frac{\theta}{2}
+2{\tt cos}^2\frac{\theta}{2} \quad .
\end{eqnarray*}
The amplitude polarization density matrix (38) can also be represented 
in the standard form
\begin{eqnarray}
\rho= \frac{1}{2} \left( \begin{array}{cc}
1+k\,{\tt cos}\theta & k\,{\tt sin}\theta e^{i\varphi} \\
k\,{\tt sin}\theta e^{-i\varphi} & 1-k\,{\tt cos}\theta 
\end{array} \right) = \frac{I}{2}+{\bf P}\,{\bf \hat S} \\ \nonumber
=\frac{1}{2}[I + k\,( 
{\tt sin}\theta{\tt cos}\varphi\, \sigma_1 +{\tt
sin}\theta{\tt sin}\varphi\, \sigma_2
+{\tt cos}\theta\, \sigma_3 )] \; ,
\end{eqnarray}
where $\bf \hat S=\{\sigma_1,\sigma_2,\sigma_3\}$ is the spin operator.
The polarization vector $\bf{P}$ is collinear to
the $b$-quark momentum and the absolute value of the polarization vector,
also
called the polarization degree, is defined by the matrix element 
of the $t\to W^+ b$ decay. In the SM  
\begin{eqnarray}
k=\frac{m_t^2-2\,m_W^2}{m_t^2+2\,m_W^2} \; = 0.41 \quad . \nonumber
\end{eqnarray}
In the case of a general interaction lagrangian the 
polarization degree depends on $f_{2L,R}$ (see (16) in sect.3.1).


\begin{thebibliography}{99}

\bibitem{peccei}
R.D. Peccei, X. Zhang, Nucl.Phys. {\bf B337}, 269 (1990) \\
R.D. Peccei, S. Peris, X. Zhang,  Nucl.Phys. {\bf B349}, 305 (1991)

\bibitem{boos1}
E. Boos, A. Pukhov, M. Sachwitz, H.J. Schreiber,
Z.Phys. {\bf C75}, 237 (1997); Phys.Lett. {\bf B404}, 119 (1997).

\bibitem{cao}
J.-J. Cao, J.-X.Wang, J.-M. Yang, B.-L. Young, X.Zhang,
Phys.Rev. D58 (1998) 094004.

\bibitem{boos2}
E.Boos, L.Dudko, T.Ohl, Eur.Phys.J. {\bf C11} (1999) 473.

\bibitem{buchmueller-hagiwara-gounaris1}
W.~Buchm\"{u}ller, D.~Wyler, Nucl. Phys. {\bf B268},  621 (1986);\\
 K.~Hagiwara, S.~Ishihara, R.~Szalapski, D.~Zeppenfeld,
Phys. Rev. {\bf D48},  2182 (1993); \\
 K.~Hagiwara, R.~Szalapski, D.~Zeppenfeld,
Phys. Lett. {\bf B318},  155 (1993); \\
B.~Grzadkowski, J.~Wudka,
Phys. Lett. {\bf B364}, 49 (1995);  \\
G.J.~Gounaris, F.M.~Renard, N.D.~Vlachos,
Nucl. Phys. {\bf B459},  51 (1996).

\bibitem{whisnant}
K.~Whisnant, J.M.~Yang, B.-L.~Young, X.~Zhang,
Phys. Rev. {\bf D56}, 467 (1997).

\bibitem{KLY}
G.L.~Kane, G.A.~Ladinsky, C.-P.~Yuan, Phys.Rev. {\bf D45} (1992) 124.

\bibitem{cleo}
M.~Alam et al., CLEO Collaboration, Phys. Rev. Lett. {\bf 74}, 2885
(1995).
 
\bibitem{larios}
L.~Larios, M.A.~Perez, C.-P.~Yuan,
Phys.Lett. {\bf B457} (1999) 334.

\bibitem{caso}
C.~Caso et al.,
Particle Data Group, Eur.Phys.J. {\bf C3} (1998) 1.

\bibitem{hikasa}
K.Hikasa, K.Whisnant, J.Yang, B.Young, Phys.Rev. {\bf D58} (1998) 114003 

\bibitem{CompHEP}
E.~Boos, M.~Dubinin, V.~Ilyin, A.~Pukhov, V.~Savrin, preprint INP MSU
94-36/358, 1994, hep-ph/9503280; \\
P.Baikov et.al, in: {\it Proc.of X Workshop on High Energy Physics and
Quantum Field
Theory}, ed.by B.Levtchenko, V.Savrin, Moscow, 1996, p.101; \\
A.~Pukhov et.al., hep-ph/9908288; \\
see also {\tt http://theory.npi.msu.su/\~{}comphep}.

\bibitem{CDDKZ}
S.Y.~Choi, A.~Djouadi, H.~Dreiner, J.~Kalinowski, P.M.~Zerwas, 
Eur. Phys. J. {\bf C7} (1999) 123.

\bibitem{BCK}
M.S.~Baek, S.Y.~Choi, C.S.~Kim, Phys.Rev. {\bf D56} (1997) 6835.


\bibitem{KRZ}
J.H.~K\"{u}hn, A.~Reiter, P.M.~Zerwas, Nucl.Phys. {\bf B272} (1986) 560. 

\bibitem{AS}
T.~Arens, L.M.~Sehgal, Nucl.Phys. {\bf B393} (1993) 46;\\
R.H.~Dalitz, G.R.~Goldstein, Phys.Rev {\bf D45} (1992) 1531.

\bibitem{LY}
G.A.~Ladinsky, C.-P.~Yuan, Phys.Rev. {\bf D49} (1994) 4415.

\bibitem{KPR}
G.L.~Kane, J.~Pumplin, W.~Repko, Phys.Rev.Lett. {\bf 41} (1978) 1689.

\bibitem{spincorr}
G.~Mahlon, S.~Parke, Phys.Rev. {\bf D53} (1996) 4886; \\
S.~Parke, Y.~Shadmi, Phys.Lett. {\bf B387} (1996) 199.

\bibitem{topdecay}
M.~Jezabek, J.H.~K\"{u}hn, Phys.Lett. {\bf B329} (1994) 317.

\bibitem{BFU}
A.~Brandenburg, M.~Flesch, P.~Uwer, Phys.Rev. {\bf D59} (1999) 014001.

\bibitem{HZ}
K.~Hagiwara, D.~Zeppenfeld, Nucl.Phys. {\bf B274} (1986) 1.

\bibitem{B}
R.Brinkmann, preprint TESLA 99-15, to appear in: {\it Proc. of the
Workshop on Physics and Experiments with Linear Colliders},
Sitges, Barcelona, 1999.
\end{thebibliography}
\end{document}